\documentclass[pra,twocolumn,superscriptaddress,amsmath,amssymb,aps,nofootinbib]{revtex4-2}
\usepackage[utf8]{inputenc}
\usepackage{amsmath,amsthm,amssymb,amsfonts,braket}
\usepackage{placeins}[verbose]
\usepackage{tabularx, longtable, multirow}
\usepackage{graphicx}
\usepackage{dcolumn}
\usepackage{bm,dsfont}
\usepackage[english]{babel}
\usepackage{comment}
\usepackage[dvipsnames]{xcolor}
\usepackage{siunitx}
\usepackage[normalem]{ulem}
\usepackage[T1]{fontenc}

\def\br{{\mathbf{r}}}
\def\bq{{\mathbf{q}}}

\def\bk{{\mathbf{k}}}
\def\bxi{{\boldsymbol \xi}}
\def\bkappa{{\boldsymbol \kappa}}

\begin{document}
	
\title{Neutron spin echo is a ``quantum tale of two paths''}

\author{S. McKay}
\affiliation{Department of Physics, Indiana University, Bloomington IN 47405, USA}
\affiliation{Center for Exploration of Energy and Matter, Indiana University, Bloomington, 47408, USA}
\author{A. A. M. Irfan}
\thanks{Work performed at (1), current address (3)}
\affiliation{Department of Physics, Indiana University, Bloomington IN 47405, USA}
\affiliation{Institute for Quantum Computing, University of Waterloo, Waterloo, N2L 3G1, ON, Canada}
\author{Q. Le Thien}
\affiliation{Department of Physics, Indiana University, Bloomington IN 47405, USA}
\author{N. Geerits}
\affiliation{Atominstitut, TU Wien, Stadionallee 2, 1020 Vienna, Austria}
\author{S. R. Parnell}
\affiliation{Faculty of Applied Sciences, Delft University of Technology, Mekelweg 15, 2629 JB Delft, The Netherlands}
\author{R. M. Dalgliesh}
\affiliation{ISIS, Rutherford Appleton Laboratory, Chilton, Oxfordshire, OX11 0QX, UK}
\author{N. V. Lavrik}
\affiliation{Center for Nanophase Materials Science, Oak Ridge National Laboratory, Oak Ridge, TN 37831, USA}
\author{I. I. Kravchenko}
\affiliation{Center for Nanophase Materials Science, Oak Ridge National Laboratory, Oak Ridge, TN 37831, USA}
\author{G. Ortiz}
\affiliation{Department of Physics, Indiana University, Bloomington IN 47405, USA}
\affiliation{Quantum Science and Engineering Center, Indiana University, Bloomington, IN 47408, USA}
\affiliation{Institute for Quantum Computing, University of Waterloo, Waterloo, N2L 3G1, ON, Canada}
\author{R. Pynn}
\email{rpynn@iu.edu}
\affiliation{Department of Physics, Indiana University, Bloomington IN 47405, USA}
\affiliation{Center for Exploration of Energy and Matter, Indiana University, Bloomington, 47408, USA}
\affiliation{Quantum Science and Engineering Center, Indiana University, Bloomington, IN 47408, USA}
\affiliation{Neutron Sciences Directorate, Oak Ridge National Laboratory, Oak Ridge, TN, 37830, USA}

\date{\today}
	
\begin{abstract}
We describe an experiment that strongly supports a two-path interferometric model in which the spin-up and spin-down components of each neutron propagate coherently along spatially separated parallel paths in a typical neutron spin echo small angle scattering (SESANS) experiment.
Specifically, we show that the usual semi-classical, single-path treatment of Larmor precession of a polarized neutron in an external magnetic field predicts a damping as a function of the spin echo length of the SESANS signal obtained with a periodic phase grating when the transverse width of the neutron wave packet is finite. 
However, no such damping is observed experimentally, implying either that the Larmor model is incorrect or that the transverse extent of the wave packet is very large.
In contrast, we demonstrate theoretically that a quantum-mechanical interferometric model in which the two mode-entangled (i.e. intraparticle entangled) spin states of a single neutron are separated in space when they interact with the grating accurately predicts the measured SESANS signal, which is independent of the wave packet width.
\end{abstract}
	
\maketitle
\section{Introduction}

\begin{figure*}[ht]
    \centering
    \includegraphics[width=\linewidth]{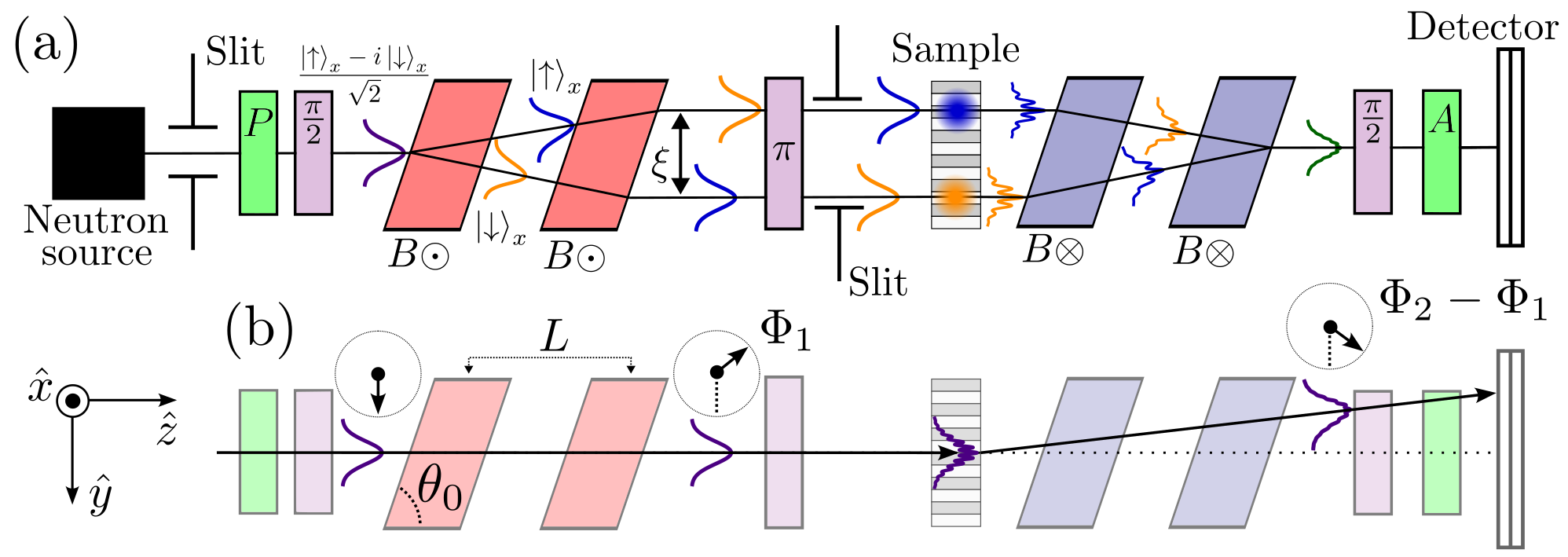}
    \caption{\label{fig:experiment setup} Simplified diagrams of a typical SESANS setup using rf flippers. (a) The two-path quantum model (see Sec. \ref{sec:quantum}) in which the two rf flippers before the sample separate the initial neutron wave packet (purple) into two mode-entangled (i.e. intraparticle entangled) wave packets (orange and blue) that are spatially separated by the spin echo length $\xi$. After interacting with the phase grating, the last two rf flippers spatially recombine the two scattered wave packets (green). The final $\pi/2$ flipper, polarization analyzer, and detector then measure the contribution of the overlapped states to the neutron polarization along the $\hat{x}$-direction.
    (b) The single-path Larmor precession model (see Sec. \ref{sec:Larmor precession}) wherein the wave nature of the neutron is treated separately from its spin: the polarization of the scattered beam arises from the difference in Larmor precession phase before ($\Phi_1$) and after ($\Phi_2$) the sample.
    The scattering angle, wave packet separation, and intrinsic coherence width are all greatly exaggerated for clarity.}
\end{figure*}

In 1969, Shull \cite{Shull_1969} determined experimentally a lower limit for the spatial extent of the coherent wavefront of a neutron wave packet using diffraction from a single slit. Because the wavelength of the neutrons used in the experiment was small (about 0.5 nm) compared to the slit width (up to 20 $\si{\micro\meter}$), the neutrons were diffracted through very small angles of about 50 $\si{\micro\radian}$.
To measure such small angles, Shull placed the slit between two perfect silicon crystals whose diffracting planes were parallel to one another. As the second crystal was rotated, it reflected neutrons over a very narrow angular range defined by its Darwin width. The increase in beam divergence caused by the single-slit diffraction becomes measurable in this case, provided the slit is not too wide. Shull made measurements with slits of three different widths and was able to observe diffraction broadening for each of them. The largest slit had a width of 21 $\si{\micro\meter}$, which Shull thus identified as a lower bound for the width of the coherent wavefront of the neutrons used. 

Subsequent measurements of single-slit diffraction made by Gahler and Zeilinger \cite{Gahler_Zeilinger} using neutrons with wavelengths between 1.5-3.0 nm confirmed Shull’s results and increased the lower bound for the coherence width at these wavelengths to around 90 $\si{\micro\meter}$. Rather than using perfect crystals as Shull had, Gahler and Zeilinger used narrow slits to collimate their incident neutron beam and a long flight path and position-sensitive neutron detector to record neutrons as a function of scattering angle. More recently, using an ultra-collimated neutron beam, Wagh {\it et al.} \cite{Wagh} measured scattering from a diffraction grating with a period of 200 $\si{\micro\meter}$, deducing a coherence width of 175 $\si{\micro\meter}$. Using a completely different method that involved neutron reflection from a diffraction grating, Majkrzak and his collaborators found values of the coherence width close to 50 $\si{\micro\meter}$ \cite{Majkrzak_2022}.
As explained in detail by Majkrzak {\it et al.}, the intrinsic coherence width of a single neutron differs from the transverse coherence width of the neutron beam, as measured, for example, by Pushin {\it et al.} \cite{Pushin_2008}.

All of the measurements of the coherence width that use single slits or diffraction gratings in transmission geometry require accurate measurements of the intensity of neutrons scattered through very small angles. The wider the diffracting slit or the greater the grating period, the more accurately the angular deviations of the neutron paths needs to be measured, implying the need for smaller Darwin widths or enhanced beam collimation, resulting in a concomitant decrease in neutron intensity. Ultimately, such measurements are thus limited for wide slits by the available neutron flux, setting an upper limit on the value of the coherence width that is measurable by these conventional methods. 

There is another method, now increasingly used to probe large structures, that makes use of neutron spin echo (NSE) invented by Mezei \cite{Mezei_1972} to measure very small angle scattering.  With this technique, called \textit{Spin Echo Small Angle Neutron Scattering} (SESANS) \cite{Rekvedlt_1996}, individual neutron magnetic moments precess in magnetic field regions through an angle that depends on their trajectory and wavelength. Each scattered neutron contributes to the spin echo polarization an amount equal to the cosine of the difference in the precession angles before and after the scattering event. The scattering angle is therefore ``labeled'' by the neutron spin, a process called \textit{Larmor labeling}; the scattering angle can be recovered using neutron polarization analysis.
SESANS is able to resolve scattering angles in the $\si{\micro\radian}$ range.
When an integrating detector is used, the experimentally measured echo polarization of the scattered neutrons is given, in general, by 
\cite{Rekveldt_2005}
\begin{equation} \label{eq:semi-classical echo pol}
    P(\xi) = \frac{\int_{-\infty}^{\infty} dq_x \int_{-\infty}^{\infty} dq_y \, \frac{d \sigma}{d \Omega}(\bq) \cos(q_y \xi)}{\int_{-\infty}^{\infty} dq_x \int_{-\infty}^{\infty} dq_y  \, \frac{d \sigma}{d \Omega}(\bq)} ,
\end{equation}
where $\xi$ is the spin echo length (also called the entanglement length \cite{Shen_2020}), $\bq~=~(q_x,q_y,q_z)$ the momentum transfer, and $d \sigma /d \Omega$ the differential scattering cross section of the sample. Equation~\eqref{eq:semi-classical echo pol} assumes that the neutron beam is incident in the $z$-direction and the scattering occurs in the $x$-$y$ plane (see Fig. \ref{fig:experiment setup}).
The length $\xi$ can be thought of as the real-space distance probed in the sample.
For the SESANS configuration at the Larmor beamline at the ISIS neutron source which uses radio-frequency (rf) neutron spin flippers, the spin echo length is given by
\begin{equation} \label{eq:spin echo constant}
    \xi = \frac{2 m f L \cot{\theta_0}}{h} \lambda^2 = \xi_0 \lambda^2,
\end{equation}
where $m$ is the mass of the neutron, $h$ Planck's constant, $f$ the (linear) frequency of the rf flippers, $\lambda$ the wavelength of the neutron, $L$ the distance between the rf flippers in each arm, and $\theta_0$ the angle between the field boundary and the optic axis. We will refer to the constant $\xi_0$ as the \textit{spin echo constant}. 
Also, because the spin echo signal is only sensitive to scattering in the $y$-direction, we will refer to this direction as the \textit{encoding direction}.
Since the SESANS method is relatively insensitive to the divergence of the neutron beam, it does not suffer from the same intensity limitations as traditional scattering methods that measure small scattering angles using extreme beam collimation.
We will call the model described above the \textit{semi-classical model} of neutron propagation, as the neutron spin precession in the magnetic field is treated independently of the neutron's wave nature.

In this model, it seems obvious that one should be able to measure the coherence width of the neutron wavefront using SESANS with a transmission grating similar to the one employed by Treimer {\it et al.} \cite{Treimer_2006} in their Ultra-Small Angle Neutron Scattering (USANS) experiments. In fact, simple calculations based on Eq.~\eqref{eq:semi-classical echo pol} described in Sec. \ref{sec:Larmor precession} indicate that the amplitude of the peaks in the SESANS pattern obtained with a transmission grating should \textit{decrease} with increasing spin echo length due to the finite coherence width of the neutron wavefront.

As we report here in Sec. \ref{sec:experment}, this is \textit{not} what we observe. Rather, the damping of the measured spin echo peaks can be accounted for by known (small) contributions to the experimental resolution arising primarily from the neutron wavelength spread and neutron beam divergence. Once these effects are removed, the echo pattern is the same as would be obtained if each neutron behaved as a perfect plane wave; there is no observed damping of the spin echo peaks.
As we show, understanding this result requires us to calculate the SESANS signal based on the ``quantum'' picture of SESANS in which the spin-up and spin-down components of the neutron wave function are spatially separated such that each state interacts coherently with a different part of the grating sample as depicted in Fig.~\ref{fig:experiment setup}(a). This fully quantum mechanical calculation presented in  Sec. \ref{sec:quantum} predicts that the SESANS pattern should indeed be \textit{independent} of the shape and width of the neutron wave packet.
Thus, the real-space picture of NSE, first described by Gahler {\it et al.} \cite{Gahler_Golub_1996,Gahler_1998}, is not simply an alternative way of describing NSE: it is an indispensable first step in a correct description of SESANS, and of NSE more generally \cite{Mezei_1988}.
We note that this interferometric model and the experimental results presented in this work are totally consistent with our previous contextuality experiments \cite{Shen_2020,Kuhn_2021} where we demonstrated the entangled nature of the neutron beam.

\section{Expectations from the single-path Larmor precession theory} \label{sec:Larmor precession}

The measurements performed using conventional neutron scattering methods have confirmed that a diffraction grating yields an increase in the divergence of a neutron beam. The experiment by Treimer {\it et al.} \cite{Treimer_2006} using a double crystal diffractometer perfectly illustrates this point for a silicon grating with rectangular, 70 $\si{\micro\meter}$ deep channels separated from each other by 16 $\si{\micro\meter}$. This object was illuminated by a neutron wave whose coherence function has an amplitude and a width, as depicted schematically by the Gaussian curve at the sample position in Fig. \ref{fig:experiment setup}(b).
We expect the scattering cross section to be equal to the result obtained for an infinite plane wave convolved with a $\bq$-dependent function whose width is the inverse of the real space coherence width.
There is a complete analogy between this result and the usual broadening of Bragg peaks seen in neutron powder diffraction that arises from finite grain size. In the present case, grain size is simply replaced by a coherence width which cuts off scattering contributions beyond a certain length scale.
Treimer {\it et al.} \cite{Treimer_2006} fitted their observed neutron scattering patterns to a scattering cross section $d \sigma / d \Omega$ obtained by applying this model and thereby estimated the width of the coherence function to be between 80 and 100 $\si{\micro\meter}$.

\begin{figure}[t]
    \centering
    \includegraphics[width=.95\linewidth]{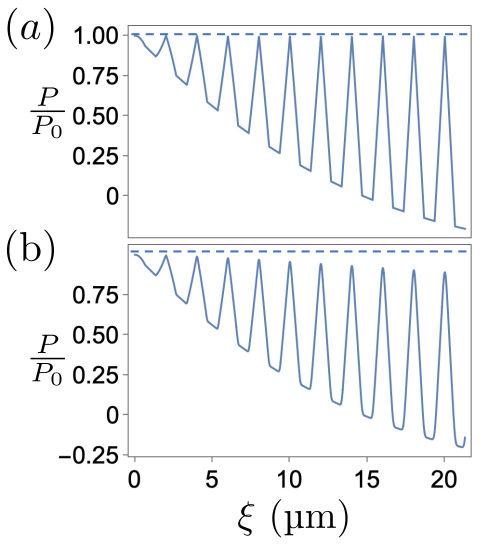}
    \caption{\label{fig:semi-classical expectation}
    (a) SESANS echo polarization $P$ normalized by the empty-beam polarization $P_0$ calculated in the single-path model for a $\SI{2}{\micro \meter}$ period silicon grating illuminated by a neutron beam incident along the normal to the plane of the  grating. The calculation used the parameters of the grating and the Larmor instrument described in the text and assumed that each neutron is an infinite plane wave. Notice that peak amplitudes return to unity for all orders.
    (b) SESANS polarization predicted for the same grating using the semi-classical model of SESANS including a finite neutron intrinsic coherence width of $\SI{60}{\micro\meter}$. The effect of a finite coherence width is to increasingly reduce the peak amplitudes as spin echo length increases and to slightly modify the sloping background. The sloping background in (b) and (c) is a result of using a pulsed neutron time-of-flight source with multiple neutron wavelengths.}
\end{figure}

A semi-classical description of a SESANS measurement using rf flippers is sketched in Fig. \ref{fig:experiment setup}(b). A neutron spin obtains a Larmor phase $\Phi_1$ before the grating sample and, after being scattered through a small angle, experiences an opposite precession through a phase angle $\Phi_2$ after the sample. The polarization of the scattered beam is then given by Eq. \eqref{eq:semi-classical echo pol} 
as described in more detail in Rekveldt's original paper \cite{Rekvedlt_1996}. 
Because the differential cross section in Eq. \eqref{eq:semi-classical echo pol} is a convolution of the infinite plane-wave result and a $\bq$-dependent coherence function, the SESANS pattern for the grating sketched shown in Fig. \ref{fig:experiment setup} is expected to be a product of the result obtained for a infinite plane wave, as plotted in Fig. \ref{fig:semi-classical expectation}(a), times a damping function, yielding a result like that shown in Fig. \ref{fig:semi-classical expectation}(b). To calculate the infinite plane-wave result shown in Fig. \ref{fig:semi-classical expectation}(a), we described the scattering from our silicon phase grating by the \textit{phase-object approximation} in which the grating imparts a phase $T(y)$ on the incoming wave packet \cite{deHaan,frank_1994}.
In the case of a one-dimensional grating, the imparted phase profile can be written as a piecewise periodic function ($b>a\ge 0$)
\begin{align}
    T(y)= \begin{cases}e^{i \phi} & -a<y \le a \\ 1 & \hspace*{0.26cm} a < y \leq b  
    \end{cases} ,
    \label{eq:PhaseGrating}
\end{align}
where $e^{i \phi}$ is the phase imparted by the silicon ``walls'' of the grating and $p=a+b$ is the period of the grating, so $T(y+p)=T(y)$. The phase-object approximation is  numerically indistinguishable from a full dynamical theory when neutrons are incident perpendicular to the grating structure \cite{Ashkar_2011}.

We now must slightly modify Eq. \eqref{eq:semi-classical echo pol} to calculate the echo polarization in the semi-classical, single-path model because Eq. \eqref{eq:semi-classical echo pol} assumes an incident plane wave neutron. First, since the grating is spatially uniform in the $x$-direction, we can eliminate the integral over $q_x$. Also, because we are no longer considering an incident plane wave but a wave packet with a finite transverse width, we must include an integral over impact parameter $y_0$ with a corresponding impact parameter distribution $\phi_{\sf B}(y_0)$.
Finally, the finite transverse width of the wave packet is included in the definition of the cross section as described below.
With these modifications, the echo polarization from Eq.~\eqref{eq:semi-classical echo pol} becomes
\begin{eqnarray}
    P_y(\xi_y)=\frac{\int d y_0 \phi_{\sf B}(y_0) \int d q_y \frac{d \sigma( q_y)}{ d \Omega} \cos (q_y  \xi_y  )}{\int d y_0 \phi_{\sf B}(y_0) \int d q_y  \frac{d \sigma( q_y)}{ d \Omega}}
    \label{Pyy} ,
\end{eqnarray} 
where $\bm \xi = (0,\xi_y,\xi_z)$ is the spin echo vector with $\xi = |\bm \xi|$, $\bq=(q_x,q_y,q_z)=\bk_{\sf 0}-\bk'$ is  the \textit{mean} momentum transfer, and $d \sigma( q_y)/ d \Omega$ the differential scattering cross section for a finite-sized  (spinless) wave packet. Here $\bk_{\sf 0}=(k_{{\sf 0}x},k_{{\sf 0}y}=0,k_{{\sf 0}z})$ is the mean initial momentum and $\bk' = (k_x',k_y',k_z')$ the scattered wave vector.

In the phase-object approximation, the outgoing scattered state for an incoming (spinless) wave packet is
\begin{eqnarray}
\label{eq:outgoing wavepacket spinless}
    \Phi_{\sf out}({\br,t})&=&     \int \!\! d\bk \,g(\bk)  \int \!\! d k_y' \, 
    e^{i\left[\mathbf k' \cdot \mathbf r-k_y y_0-\omega(k') t\right] }
    \ F (\kappa_y) ,\nonumber\\
    F (\kappa_y) &=& \frac{1}{2 \pi}
    \int dy' \ e^{i \kappa_y   y'} \  T(y'),  
\end{eqnarray}
where $\kappa_y=k_y-k_y'$ is the momentum transfer, $\hbar \omega(k)~=~\hbar^2 k^2/(2m)$ the total incident energy with wave vector $k^2=k_x^2+k_y^2+k_z^2$, and $g(\bk)$ the momentum distribution of the incident wave packet. For simplicity we assume a Gaussian wave packet for the incident neutron
\begin{equation} \label{eq:Gaussianwp}
\begin{aligned} 
    g(\bk) &=\frac{1}{(2\pi)^{3/2}}g_y(k_y)\delta(k_x-k_{{\sf 0}x})\delta(k_z-k_{{\sf 0}z}) , \\
    g_y(k_y) &= \sqrt{\frac{\Delta}{\sqrt{2\pi} }} \ e^{-\frac{\Delta^2 (k_y-k_{{\sf 0}y})^2}{4}} ,
\end{aligned}
\end{equation}
with $\Delta$ determining the spatial width of the wave packet in the $y$-direction at the time of scattering from the grating, assuming that the wave packet does not appreciably spread during the scattering process.

If one follows the usual presentation of scattering theory as discussed in more detail in \cite{Katz_1966,Karlovets_2015,Irfan_2021}, in the far-field, we can separate the free propagator in $\Phi_{\sf out}(\br,t)$
and identify $d \sigma( q_y)/ d \Omega \propto |f(\bq)|^2$, where 
$f(\bq)$ is given by 
\begin{eqnarray} 
    f(\bq)
    &=& \int d\bk \, g(\bk)  \,e^{-ik_{y} y_0} \, F(\kappa_y)    .  \label{eq:scattering amplitude pw}
\end{eqnarray} 
%
%
Notice that $|f(\bq)|$ is a function of $q_y$, $y_0$ and $\Delta$, and 
the familiar plane wave analysis is recovered in the $\Delta~\to~\infty$ limit,  where $|f(\bq)|$ becomes a function of $q_y$ only. This identification is consistent with the picture advocated in \cite{Treimer_2006}, namely the convolution of the wave packet's momentum distribution and the van Hove cross section for each plane wave component \cite{VanHove_1954}.

From the above equations, we see that the echo polarization is given by the Fourier transform of the differential cross section, which includes the wave packet size in its definition, integrated over the impact parameter $y_0$. Assuming that the transverse size of the beam is larger than $\Delta$, $p$, and $\xi_y$, one can approximate $\phi_{\sf B}(y_0)\approx 1$. In this case, the integral over impact parameter allows the result to be expressed as a product of the plane wave result and a Gaussian damping function that represents the effect of the finite width of the wave packet.
When the spin echo length is independent of the neutron wavelength, the echo polarization given in Eq. \eqref{Pyy} simplifies to
\begin{eqnarray} \label{eq: final semi py}
    P_y(\xi_y)= G(\xi_y,\Delta) \, \frac{1
    }{ p  } 
     \int_{-\frac{p}{2}}^{\frac{p}{2}} dy  \ T^*(y) \  T(y + \xi_y) ,
\end{eqnarray}
where $G(\xi_y,\Delta)$ is a damping function that depends on the spin echo length and initial wave packet transverse size. 
In particular, the damping function $G(\xi_y,\Delta)$ for our Gaussian wave packet is given by
\begin{eqnarray}
    G(\xi_y,\Delta)=  e^{-\frac{\xi_y^2}{2\Delta^2}}.
\end{eqnarray}
Again, notice that the standard plane wave result is obtained in the $\Delta \to \infty$ limit in which case the damping function is identically unity, leaving just the autocorrelation function (i.e., Patterson function) of the grating \cite{Andersson_2008} in Eq. \eqref{eq: final semi py}. 

\FloatBarrier
\section{Expectations of a two-path quantum model} \label{sec:quantum}

\begin{figure}[t]
    \centering 
    \includegraphics[width= \columnwidth]{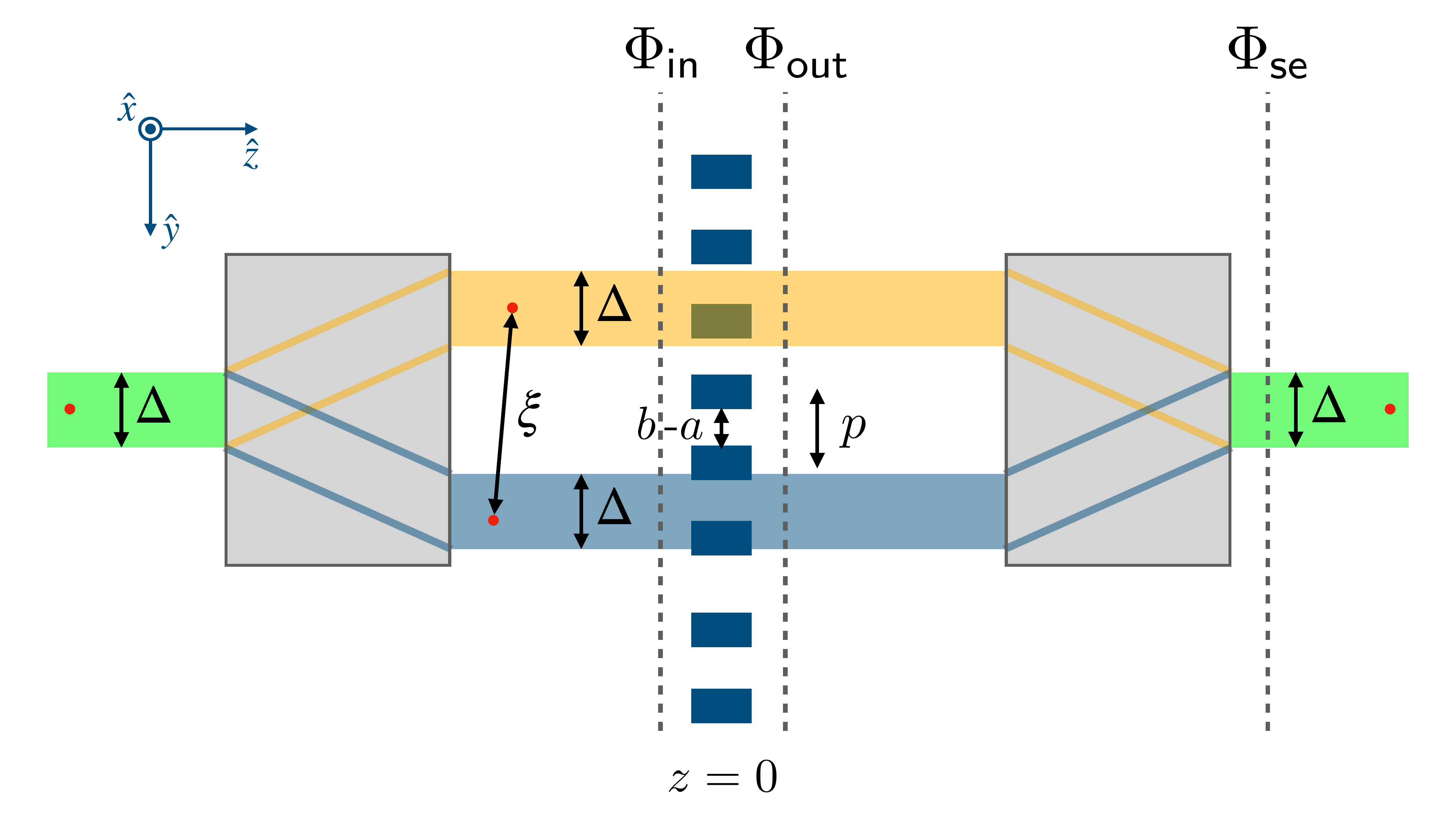}
    \caption{Diffraction of an entangled neutron state from a phase grating of period $p=a+b$ with $a$ the width of the grating walls and $b-a$ the width of the channels.
    The light (orange) and darker (blue) paths represent the spin-up and spin-down wave packet components, respectively. The transverse wave packet width and path separation are represented by $\Delta$ and $\xi$, respectively. The grating is positioned at $z=0$.}
    \label{fig: Diffraction}
\end{figure}

In the semi-classical model SESANS, each neutron follows a single well-defined path through the apparatus shown in Fig. \ref{fig:experiment setup}(b). However, this model ignores the spin-orientation-dependent refraction at the field boundaries of the flippers. In fact, a neutron whose spin is in the $y$-direction after the first $\pi/2$ flipper is a superposition of the up and down spin states in the $x$ basis; these two states will be differently refracted at the parallelogram boundaries, as indicated by the orange and blue trajectories in Fig. \ref{fig:experiment setup}(a). 
In previous work \cite{Shen_2020,Kuhn_2021}, we demonstrated the mode-entanglement of the spin and path for each single neutron passing through a SESANS apparatus, which could be robust to classical noise \cite{Wood_2014}; as we now show, this entanglement must be considered in determining the SESANS pattern obtained with a diffraction grating.

We now present the resulting polarization predicted by the scattering theory developed in \cite{Irfan_2021}. According to this theory the incoming entangled (with respect to the tensor product of path and spin distinguishable subsystem spaces) wave packet is
\begin{eqnarray} \label{eq:incoming wavepacket}
    \Phi_{\sf in}({\br,t})&=&\int 
    d \Lambda_{\mathbf k}
    \, e^{i \bk \cdot \br-i k_y\,y_0  -i \omega(k) t} \, \ket{\chi_{\bk \cdot \bxi}} ,
\end{eqnarray}
where $y_0$ is the impact parameter, $d \Lambda_{\mathbf k}=d\bk \, g(\bk)$ with $g(\bk)$ being the incident momentum distribution,
that we assume without loss of generality to be as in Eq. \eqref{eq:Gaussianwp}, 
$\hbar \omega(k)~=~\hbar^2 k^2/(2m)$ the incident energy, $k^2=k_x^2+k_y^2+k_z^2$ the total initial momentum squared, and
\begin{eqnarray}
    \ket{\chi_{\bk \cdot \bxi}}&=&\frac{e^{-i \bk \cdot \frac{\bxi}{2}}\ket{\uparrow}_x-i e^{i \bk \cdot \frac{\bxi}{2}}\ket{\downarrow}_x}{\sqrt{2}}.
\end{eqnarray}
At $t=0$, the entangled wave packet $\Phi_{\sf in}$ hits the grating located at $\br_{\sf g}=(x,y,0)$ which imparts a phase $T(y)$ [see Eq. \eqref{eq:PhaseGrating}] to the incoming wave packet, so the outgoing wave packet becomes
\begin{equation}
    \Phi_{\sf out} ( \br_{\sf g}, 0 ) = T( y ) \, \Phi_{\sf in} ( \br_{\sf g}, 0 ).
\end{equation}
Again, we use the phase-object approximation to calculate the scattering amplitude
\begin{align}
    \Phi_{\sf{out}}(\mathbf r, t) &= \!\!
    \int \!\! d \Lambda_{\mathbf k }  \int \!\! d k_y' \, 
    e^{i\left[\mathbf k' \cdot \mathbf r-k_y y_0-\omega(k') t\right] }
    \ F (\kappa_y) \ket{\chi_{\bk \cdot \bxi}}, 
\end{align}
with momentum transfer $\bkappa=\bk-\bk'$, scattered momentum
$\mathbf k'=(k_{{\sf 0}x},k'_y,k'_z)$, $k'_z=\sqrt{k_y^2+k_{{\sf 0}z}^2-{k'_y}^2}$, and $\omega(k')=\omega(k)$. Note that $\bkappa$ is different from the mean momentum transfer $\bq$, since $\bkappa$ involves momentum states distributed within a Gaussian wave packet.
Immediately after diffraction from the grating, the neutron passes through another pair of rf flippers that operates as a disentangler \cite{Lu2020}; the resulting neutron state is
\begin{align}
    \Phi_{\sf{se}}(\mathbf r, t)
    \! =& \!\!\!  \int \!\! d \Lambda_\mathbf k \ 
    \!\! \int \!\! d k_y' \, 
    e^{i\left[\mathbf k' \cdot \mathbf r-k_y y_0-\omega(k) t\right] }
    \ F (\kappa_y) \ket{\chi_{\bkappa \cdot \bxi}} .
\end{align}
We are now ready to calculate the spin polarization along the $y$ quantization axis which is defined as
\begin{eqnarray}
    \label{eq:polarization}
    P_y=\frac{\int dy_0 \phi_{\sf B}(y_0)\int dy  \, \Phi^\dagger_{\text{se}}(\mathbf r,t) \sigma^y \Phi_{\text{se}}(\mathbf r,t)   }{ \int dy_0 \phi_{\sf B}(y_0)\int dy  \, | \Phi_{\text{se}}(\mathbf r,t) |^2	},
\end{eqnarray}
where the impact parameter distribution $\phi_{\sf B}(y_0)$ represents the potentially non-uniform neutron beam profile and \textit{not} the transverse extent of the neutron wave packet.

In Eq. \eqref{eq:polarization}, numerator and denominator each involve two integrations over $k_{y}'$ that get reduced to unity after integration over $y$, implying that only plane wave components with identical $k_y'$ interfere constructively. To analytically compute the remaining $k_{y}'$ integration, we assume that $k'_z\approx k_{{\sf 0}z}$, which is valid if $k_{{\sf 0} z}\gg 1/p$, so that
\begin{multline}
\int dy \, \Phi^\dagger_{\text{se}}(\mathbf r,t) \sigma^y \Phi_{\text{se}}(\mathbf r,t) \\ 
= \int dy' \Re \big[ T^*(y'+ \frac{\xi_y}{2}) T(y'- \frac{\xi_y}{2})\big] \
    |\Phi (y' -y_0 ,t)|^2,
\end{multline}
where 
$\Phi (y,t)=e^{-i (k_{0x} x+ k_{0z}z)}\int d \Lambda_{\mathbf k} \    e^{i\bk \cdot \br+i \omega(k) t}$, and $\Re$ represents the real part. 
In addition, if the transverse size of the beam is larger than $\Delta$, $p$, and $\xi$, then one can approximate $\phi_{\sf B}(y_0)\approx 1$, so
\begin{eqnarray}
\hspace*{-0.7cm}    P_y(\xi_y)     &=&\frac{1}{p}
    \int_{-\frac{p}{2}}^{\frac{p}{2}} dy  \,   T^*(y) \, T(y + \xi_y) \ ,  
\end{eqnarray}
which turns out to be a piecewise periodic function of $\xi_y$ of period $p$, so $P_y(\xi_y+p)=P_y(\xi_y)$, which is independent of $\Delta$:
\begin{align}
P_y(\xi_y) \!&=&\!\!\begin{cases}
1-\frac{2(1-\cos \phi)}{p}\xi_y & \hspace*{0.37cm} 0 <  \xi_y \le \tilde \xi_{\sf min} \\ 
1-\frac{2\tilde\xi_{\sf min}(1-\cos \phi)}{p} & \tilde \xi_{\sf min} < \xi_y \le \tilde \xi_{\sf max}  \\
1+\frac{2(1-\cos \phi)}{p}(\xi_y-p) & \tilde \xi_{\sf max} <\xi_y \le p,
    \end{cases}
\end{align}
where $\tilde \xi_{\sf min}={\sf min}[2a, b-a]$ and $\tilde \xi_{\sf max}={\sf max}[2a, b-a]$.
Notice that even though we considered the incident neutron as a wave packet with finite transverse extent, we do not recover the damping function $G(\xi_y,\Delta)$ obtained in the semi-classical, single-path model shown in Eq. \eqref{eq: final semi py}. 
The reason is that the numerator of Eq. \eqref{eq: final semi py} includes the expression $\Phi ^* ( y'-  y_0, 0)      \Phi(   y'\pm \xi_y-  y_0  ,0)$, instead of the $|\Phi (y' -y_0 ,t)|^2$ that appears in Eq.~\eqref{eq:polarization}, whose integral over $y'$ is proportional to $G(\xi_y,\Delta)$.

The fundamental difference between the two-path and the single-path spin echo descriptions is rooted in the entangled nature of the neutron. Mathematically, the seemingly innocuous replacement $\ket{\chi_{\bk \cdot \bxi}} \rightarrow \ket{\chi_{\bk_{\sf 0} \cdot \bxi}}$ in the incoming neutron's wavefunction $\Phi_{\sf in}$ in Eq. \eqref{eq:incoming wavepacket} transforms the incoming state into an un-entangled state, and is enough to reproduce the form of the semi-classical, single-path result of the previous section.
Physically, this replacement leads to a neutron wavefunction that is a tensor product of a purely coordinate-dependent part and a purely spin-dependent part whose contribution to the echo polarization is $\bra{\chi_{\bq_1 \cdot \bxi}}\sigma^y\ket{\chi_{\bq_2 \cdot \bxi}}= \cos (\frac{\bq_1+\bq_2}{2} \cdot \bxi )$ by Eq.~\eqref{eq:polarization}, which gives the same result as in the single-path model which uses Eq. \eqref{Pyy}. 
The ``entangled'' form for the echo polarization given in Eq. \eqref{eq:polarization} 
allows for the individual components of the neutron spinor to propagate independently (either as wave packets when $\Delta~<~\infty$ or plane waves when $\Delta~\to~\infty$).

\section{SESANS Experiment with a phase grating and Analysis} \label{sec:experment}

\begin{figure*}[ht]
    \centering
    \includegraphics[width=\linewidth]{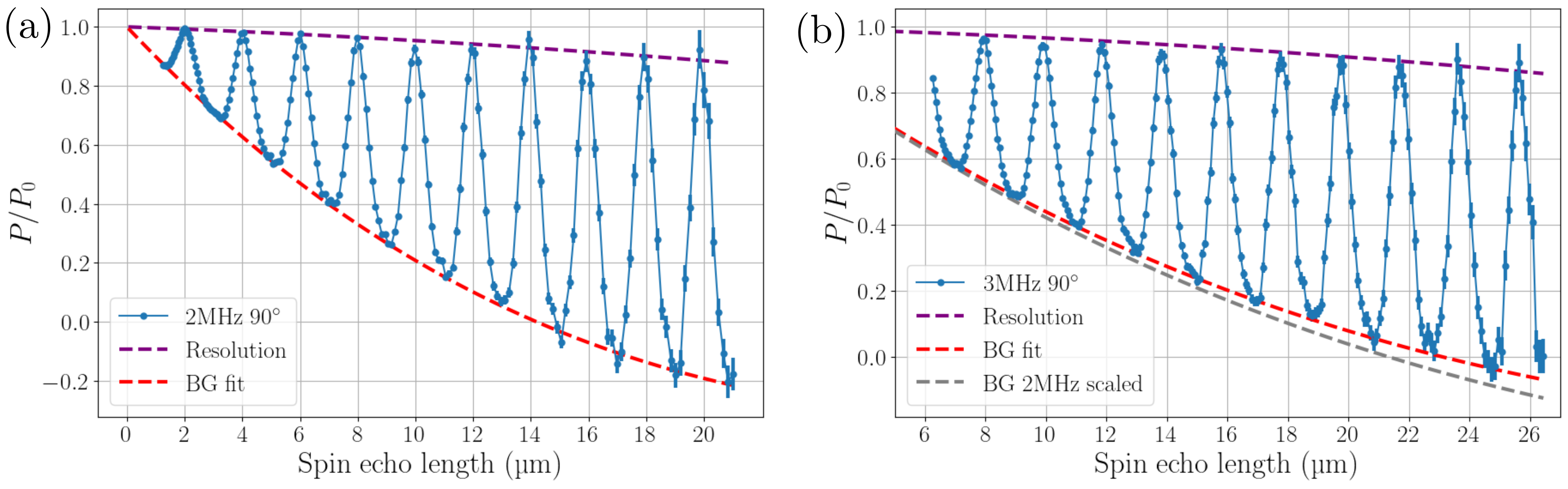}
    \caption{\label{fig:90degdata} (a) Plot of normalized spin echo polarization vs. spin echo length at an rf frequency of 2 MHz and grating angle of 90 degrees such that the grating channels were along the $\hat{x}$-direction in Fig. \ref{fig:experiment setup}. (b) The same plot for an rf frequency of 3 MHz.
    The background functions (BG), which appear because the spin echo length $\xi$ varies quadratically with the neutron wavelength $\lambda$ and the phase imparted on the neutron from the silicon grating is also proportional to $\lambda$, are fitted from the local minima of the echo polarization. 
    The resolution curves are the expected maxima of the echo polarization peaks when the theoretical echo polarization is convolved with the Gaussian instrumental resolution function [see Eq. \eqref{eq:resolution}].}
\end{figure*}

We used the Larmor instrument at the ISIS pulsed neutron source in the UK, employing rf flipper frequencies of 2 MHz and 3 MHz in separate measurements. Neutron polarization analysis was accomplished by two 1 m long polarizing supermirror v-cavities manufactured by Swiss Neutronics mounted in series. The polarizing v-shaped section in each cavity is 650 mm long and double-side coated with $m=5$ Fe/Si supermirror; the polarizer covers a beam cross section of 30$\times$30 mm inside an $m=3$ guide.
Neutrons were detected using a linear wavelength-shifting fibre detector manufactured at ISIS with 64 pixels; the detector fibers were 35$\times$0.5 mm mounted on a pitch of 0.65 mm behind a continuous sheet of Lithium scintillator glass.  The detector efficiency was about 70\% for wavelengths greater than 0.5 nm. Two slits separated by 5 m, the first being $14 \times 14 \; \si{\milli\meter^2}$ in size and the second $4 \times 4 \; \si{\milli\meter^2}$, defined the neutron beam incident on our diffraction grating; the grating's normal was kept parallel to the incident neutron beam for all measurements.
The second slit was approximately 50 mm in front of the diffraction grating sample.

The diffraction grating used in our experiment was fabricated using electron beam lithography and wafer processing at the Center for Nanophase Materials Sciences at Oak Ridge National Laboratory. Double side polished single crystal silicon wafers with (110) orientation served as a starting material for channel fabrication.  The wafers were coated with 80 nm of silicon nitride using low pressure chemical vapor deposition (LPCVD). The SiN side of the wafers was then patterned using electron beam lithography (JEOL JBX-9300 system). The lithographic pattern was aligned with respect to the wafer so that channel length was along the $\langle 112 \rangle$ direction.
After exposing and developing positive tone electron beam resist (300 nm of ZEP 520A), the pattern was transferred into the silicon nitride layer using anisotropic reactive ion etching in a C4F8 plasma. As a result of a timed etch in a hot (65$^{\circ}$C) KOH solution (30\% in deionized water), channels with rectangular cross sections were formed in regions of the silicon substrate that were not masked by the silicon nitride layer. The geometry of the resulting channels was analyzed using scanning electron microscopy (FEI NOVA 500 series) and yielded a period of 2~\textmu m, a channel depth of approximately 10 \textmu m and a channel width of approximately 560 nm. Because the electron beam lithography is controlled by a laser interferometer, the positioning of the top of each grating channel is expected to be accurate to within 10 nm.

Although electron microscopy can only provide information about the local dimensions of the features of the diffraction grating, previous neutron measurements have verified that the channel depths are very uniform over macroscopic, millimeter-sized regions of the grating. Furthermore, the SESANS pattern obtained with the grating is well described by a dynamical scattering theory that we previously showed is equivalent to the phase-object approximation assumed in this paper when the normal to the grating is aligned with the neutron beam \cite{Ashkar_2011}. 

The SESANS echo polarization $P$, normalized to the echo polarization without a sample $P_0$, was measured at rf frequencies of 2 MHz and 3 MHz with the grating channels perpendicular to the SESANS encoding direction. These data are shown in blue in Fig. \ref{fig:90degdata}. The periodic, almost triangular peaks in these plots ride on a sloping background, which is an artifact of the use of a neutron time-of-flight instrument that arises because the spin echo length $\xi$ varies quadratically with the neutron wavelength $\lambda$ [see Eq. \eqref{eq:spin echo constant}], and the phase that each neutron accumulates as it passes through the silicon grating is also proportional to $\lambda$. 
Direct calculation shows that for a grating with perfectly rectangular channels, the background below the peaks is described by the function
\begin{equation} \label{eq:background function}
    \mathrm{BG}(\xi) = 1 + (b - a) \left[\cos \left(\rho h \sqrt{\frac{\xi}{\xi_0}} \right) - 1 \right],
\end{equation}
where $(b - a)$ is the width of the grating channels and $\rho h \sqrt{\xi / \xi_0}$ is the neutron phase accumulated when passing through a grating wall of height $h$ between channels.
Here $\rho~=~\num{2.06e-4} \, \si{\nano\meter^{-2}}$ is the scattering length density of silicon. 
In practice, the spin echo constant $\xi_0$ defined in Eq. \eqref{eq:spin echo constant} was found by fitting the positions of the spin echo peaks obtained with a 1 $\si{\micro\meter}$ period silicon grating that is otherwise similar to the grating that we measured in this experiment.

It is immediately evident in Fig. \ref{fig:90degdata} that the maximum value of $P/P_0$ for spin echo peaks with a period of $\SI{2}{\micro \meter}$ decreases as the spin echo length increases, apparently in agreement with the predictions of the semi-classical description of SESANS. However, this conclusion ignores several experimental effects that tend to damp the spin echo peaks. First, the time duration of the neutron pulse emitted by the ISIS pulsed neutron source has been previously measured for Larmor and found to depend linearly on neutron wavelength. The finite neutron pulse duration implies that neutrons of slightly different wavelengths within the pulse will reach the neutron detector at the same time.
Also, even though SESANS is much less sensitive to the angular divergence of the neutron beam than most conventional neutron methods, the spin echo length has a weak dependence on the divergence.
Finally, the time-of-flight (tof) wavelength bin width (usually set to either 0.0025 nm or 0.005 nm) and the expected slight ($\lesssim \SI{10}{\nano\meter}$) misplacement of the channels arising from grating fabrication also make small contributions to the uncertainty in spin echo length.
Combining these factors, we have
\begin{equation} \label{eq:resolution}
    \frac{\delta_{\xi}^2}{\xi^2} = \frac{4 \delta_{\theta}^2}{\sin^2(2 \theta_0)} +  \frac{4 \left( \delta_{\lambda}^2 + \delta_b^2 \right)}{\xi / \xi_0} + \frac{\delta_J^2}{\xi^2}
\end{equation}
where $\delta_{\theta} = \SI{0.75}{\milli\radian}$ is the standard deviation of the beam divergence, $\delta_{J} = \SI{10}{\nano\meter}$ the misplacement of the grating channels, and $\delta_b = \SI{1e-3}{\nano\meter}$ is the standard deviation in wavelength due to the tof bin width.
To a good approximation, the wavelength uncertainty is linear in the wavelength range used in this work (0.3 - 1.3 $\unit{\nano\meter}$):
\begin{equation}
    \delta_{\lambda} = a_{\lambda} + b_{\lambda} \sqrt{\xi / \xi_0}
\end{equation}
with the fitted parameters $a_{\lambda}~=~\SI{3.33e-4}{\nano\meter}$ and $b_{\lambda}~=~\SI{1.01e-4}{}$. For simplicity, a Gaussian instrumental resolution function was assumed. 

As described above, Fig. \ref{fig:semi-classical expectation}(a) shows calculations of the spin echo pattern based on an infinite plane wave model. When this pattern is convolved with the Gaussian instrumental resolution function, the peaks are damped and there is a slight rounding at the base of each SESANS peak.
We note that each neutron pulse has weak long-wavelength tail that we did not include in the calculation and that this further increases the rounding in the valleys between the SESANS peaks but has negligible effect on the peak heights. The normalized peak echo polarizations obtained by convolving the plane wave pattern with the instrumental resolution are plotted as purple dashed lines labeled ``Resolution'' in Fig. \ref{fig:90degdata}.
Evidently, the instrumental resolution explains quantitatively the observed departure of the maximum value of $P/P_0$ from unity for spin echo lengths up to 25 \textmu m or more: within the statistical uncertainty of our measurements, there is no residual damping of the SESANS peaks that could be attributed to finite coherence width.

Of course, we would also expect the instrumental resolution to slightly broaden the spin echo peaks at the same time as it decreases their amplitudes. While this effect does account for most of the increase in the fitted peak widths with spin echo length, it does not account for the fact that the lowest order peak is broader than the the simple plane wave calculation predicts.
Most of this discrepancy can be explained by the fact that the phase contrast of the grating departs from a pure rectangular profile and is somewhat trapezoidal. While some part of the trapezoidal shape could result from grating fabrication, most is due to the beam divergence: neutrons that do not travel exactly perpendicular to the plane of the grating will contribute to a trapezoidal phase profile \cite{Ashkar_2011}. In addition to slightly broadening the spin echo peaks, this effect also causes more rounding of the valleys between the peaks. However, it does not cause any further decrease of the peak heights provided that all of the walls between grating channels are identical.

\begin{figure}[t]
    \centering
    \includegraphics[width=\linewidth]{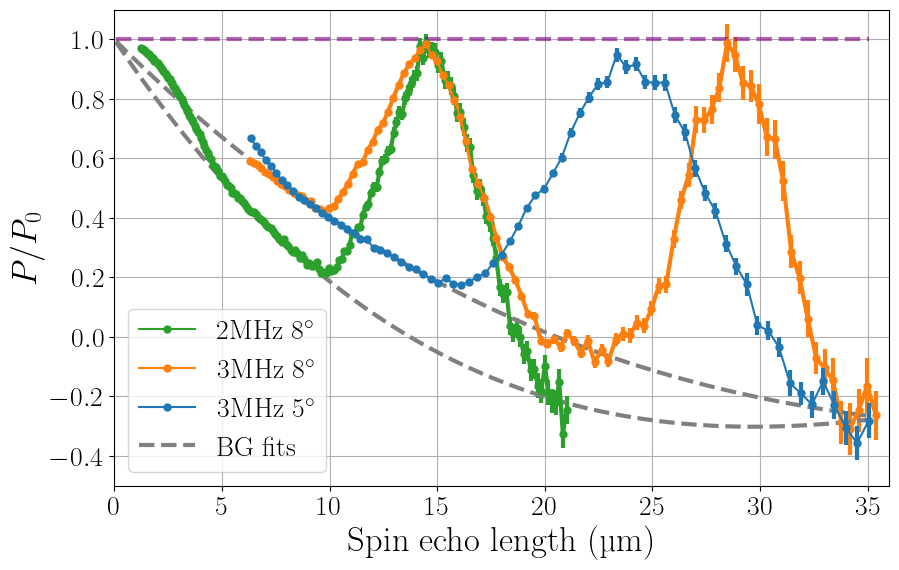}
    \caption{\label{fig:low-angle data} Plot of normalized spin echo polarization vs. spin echo length at an rf frequency of 2 and 3 MHz and grating angles of 8 and 5 degrees. The gray dashed lines are the background (BG) fits as described in Sec. \ref{sec:experment} and in the caption of Fig. \ref{fig:90degdata}. Notice that the amplitude of the peaks return to unity (purple dashed line).}
\end{figure}

To further confirm our conclusion that the intrinsic coherence width of the neutron wave packet does not influence the height of spin echo peaks, we performed experiments in which the grating was rotated about the optic axis such that the grating channels were inclined at either 8$^{\circ}$ or 5$^{\circ}$ to the spin echo encoding direction. This effectively multiplies both the period of the grating and the channel width by the same factor.
In this case, the fitted background function changes very slightly, as shown in Fig. \ref{fig:low-angle data}, and the spin echo peak is much broader. As expected, both the widths and mean spin echo lengths of these peaks are larger than the values for the corresponding peaks in Fig. \ref{fig:90degdata} by the same multiple.
Because the spin echo peaks shown in Fig. \ref{fig:low-angle data} are much broader than the instrumental resolution, the latter has no perceptible effect on the height of the $P/P_0$ peaks which remain at unity within statistical error.
If the decreasing height of the spin echo peaks were due to a finite coherence width, we would expect the peak height to depend only on the spin echo length at which it is observed, meaning that the peaks in Fig. \ref{fig:low-angle data} should have the same height as the peaks in Fig. \ref{fig:90degdata} at the same spin echo length, which is clearly not the case. 

\section{Discussion and Conclusion}

In conclusion, our experimental results are consistent with the predictions of a two-path interferometric model of a mode-entangled neutron beam, which predicts that the SESANS signal is independent of the wave packet width.
The known contributions to instrumental resolution (principally the wavelength spread due to the finite neutron pulse width and beam divergence) explain all of the observed damping of the SESANS signal. While the statistical uncertainty of the data prevent us from completely falsifying the single-path Larmor precession model, the data are incompatible with a single-path model in which the wave packet has a full-width half-maximum of less than about $\SI{150}{\micro \meter}$.  
However, in the two-path model, the two neutron spin states are each represented by wave packets and follow paths that are separated at the sample position by the spin echo length. Therefore, each state visits a different part of the scattering sample, as shown in Fig. \ref{fig:experiment setup}(a), and has its phase modified by the sample. The SESANS apparatus brings the two wave packets into spatial overlap and projects out a component of the neutron polarization. The coherence width of the neutron plays no role precisely because the two spin states are brought back into nearly perfect overlap.

Once overlapped, each spatial coordinate within one packet coincides with a coordinate in the other packet that was separated from it by the spin echo length when the two states interacted with the sample.
Integration of the neutron polarization over the wave packet spatial extent thus separates into a product of an integration over the packet and an integration over the sample of the same function that applies for scattering by a perfect plane wave. Even if the wave packet were very small, the different impact parameters of neutrons within the neutron beam would lead to the same effect, except in the case where the neutron beam width was less than the period of the scattering grating.
It is interesting to note that the model describing the SESANS experiment is essentially the same as that used to describe differential interference contrast microscopy, with the main difference being that the latter method uses focusing optics to create a complete image while, in SESANS, integrating over the neutron detector produces a two-point density correlation function.

As an aside, the expressions in Sec. \ref{sec:quantum} that describe the two-path model are not quite correct because the spin echo length in the scattered beam depends on $\bq$ and is only identical to the value imposed by the first two rf flippers when $\bq = 0$. Therefore, the two wave packets are not quite brought into overlap, and the departure from exact overlap depends on the scattering angle.
It is straightforward to estimate the standard deviation of the scattering angle for each wavelength and to include this extra term in the instrumental resolution. For our experiment this added term had a negligible effect: it is an order of magnitude less than the terms representing wavelength spread and beam divergence and can safely be ignored. However, for gratings with smaller periods than considered in this work, this effect might not always be the case.

The final result of our experiment and its analysis is an equation for the SESANS echo polarization that is identical to the semi-classical Larmor precession model that is conventionally used to analyze SESANS data. However, both the physical picture and mathematical theory developed in this work are different, with each neutron spinor following a different path through the apparatus rather than a single path, which requires additional semi-classical assumptions. Our two-path analysis clearly shows that the sample's autocorrelation function deduced from SESANS is indeed the same as the one obtained from van Hove's theory. Even if future experiments were to unambiguously demonstrate the finite width of a neutron wave packet, this result would not change our understanding of what SESANS measures because the wave packet width does not appear in the final expression for the echo polarization.

On the contrary, conventional neutron scattering experiments could be influenced by the finite coherence width to a degree that depends on both the scattering strength of the sample and the length scales probed.
For example, the dependence of the cross section on the finite nature of the coherence width has been experimentally demonstrated in proton-H$_2$ and helium-ion scattering experiments using an analysis similar to the one we presented in Sec. \ref{sec:Larmor precession} \cite{Egodapitiya_2011, Navarrete_2017, Schulz_2017}.

\begin{acknowledgements}
    The IU Quantum Science and Engineering Center is supported by the Office of the IU Bloomington Vice Provost for Research through its Emerging Areas of Research program. We acknowledge support from the US Department of Commerce through cooperative agreement number 70NANB15H259. This material is based upon work supported by the U.S. Department of Energy, Office of Science, Office of Workforce Development for Teachers and Scientists, Office of Science Graduate Student Research (SCGSR) program. The SCGSR program is administered by the Oak Ridge Institute for Science and Education for the DOE under contract number DE-SC0014664. Grating fabrication was part of a user project at the Center for Nanophase Materials Sciences (CNMS), which is a US Department of Energy, Office of Science User Facility at Oak Ridge National Laboratory. This research was undertaken thanks in part to funding from the Canada First Research Excellence Fund.
\end{acknowledgements}

\FloatBarrier
\bibliography{sources.bib}
\end{document}